# Construction and Adaptability Analysis of User's Feature Models Based on Check-in Data in LBSN

Yuanbang Li[1]   Shi Dong[1]   Chi Xu[1]

[1] Zhoukou Normal University, East Wenchang Street, Chuanhui District, Zhoukou City 466001, Henan Province,China

**Abstract.** With the widespread use of mobile phones, users can share their location anytime, anywhere, as a form of check-in data. The user context involved in these data and analysts' view towards these data are diverse. These data reflect user features, furthermore, the features rules for different users vary. Therefore, how to analyze and quantify the impact of user context on different analyst views, how to discover a user's feature from their related data and how to validate whether a feature model is suited to a user is of great significance for providing users with contextualized services in complex mobile applications. In this study, firstly, the influence of user context to analysts' view is quantitative analyzed. Secondly, multiple feature models from different views for each user are constructed. Thirdly, whether a feature model is applicable to a user is validated. Fourthly, a unified model, muti-channel convolutional neural network (CNN) is used to characterize this applicability. And finally, three data sets from multiple sources are used to verify the validity of the method, the results of which show the effectiveness of the method.

## 1 Introduction

Owing to the maturity and diversification of mobile application services, mobile Internet is integrating into people's daily lives and changing our work and learning environments. According to the 45th Statistical Report on the Development of China's Internet in 2020 by the China Internet Information Center [1], as of March of 2020, the number of Chinese netizens reached 904 million, of which 897 million were mobile netizens, accounting for 99.3%. With the widespread use of mobile phones with built-in GPS, location-based social networks (LBSNs)[2] have achieved a rapid development because users can share their physical position and send various types of information and comments freely through such networks.

Different users may have different features for various roles and backgrounds. The features of the same user may be different in different scenarios [3–4]. Various approaches to modeling user features have recently been constructed. Some researchers, including Cheng, have concentrated on the temporal features of users, focusing on the temporal relation in user check-in data and using a personalized Markov chain to model temporal user features and calculate the probability that one user will check-in at a certain location at a specific time [5]. In addition, Xu built a categorical-temporal distribution feature model of points of interest (POIs) for use within a 24 h period and analyzed the overall changes in popular POI categories throughout the day [6]. Other researchers have focused on geographical features. For example, Noulas revealed user activity patterns from check-in data and found that 20% of the check-in data appear within 1 km, 60% appear between 1 and 10 km, and the final 20% appear beyond 10 km [7]. In addition, Cheng found that the check-in data of a user often appear around multiple centers and proposed a multi-center Gaussian model to describe the geographical features of users [8].

A large number of studies have focused on building user feature models and recommending services to specific users using such models [9–11]; however, research on whether the established feature models are applicable to users has not attracted sufficient attention. In fact, different users are suited to different feature models. Through a questionnaire we organized for Master's Degree students in computer science, class of 2019 at Wuhan University, we found that some students show regularity in their schedules, eat at a fixed time, and take a walk after dinner every day, whereas others are irregular, although in terms of geographical location, they often do fixed things in a fixed place. These results show that a general feature model may be applicable to some users rather than to others. Building diversified feature models and analyzing the applicable user groups of such models is an urgent problem to be solved.

Based on the findings above, we aim to design a strategy to validate whether a user is suited to a feature model. The main contributions of this study are as follows:

1. Quantifying the influence of user context to analysts' views. Information entropy gain method is used to quantify the influence of user context to analysts' views, only when the influence is greater than a threshold, will it be taken into account.

2. constructing multiple feature models for users. We should describe the user's features from multiple perspectives to discover the most suitable feature. In this study, we use as many elements in the data set as possible that affect the user features in building user feature models, particularly including temporal, distance, and content.

3. Determining the different feature for different users. To analyze the applicable user groups for each of the feature models, we propose an algorithm to validate whether a user is applicable to a particular model.

4. Proposing a unified model to describe the applicability of different users to different feature models to support the provisioning of services.

Finally, experiments were conducted to verify the effectiveness of the proposed method.

The remainder of this paper is organized as follows. Related works are described in section 2. The proposed method is described in section 3. The experiments are presented in section 4. Finally, validity threat and conclusion are presented in section 5 and 6.

## 2 Related works

The extraction of user features is a hot research topic in LBSNs, and many methods have been proposed to extract user features. These methods can be divided into two categories based on the extraction patterns: explicit and implicit extraction methods.

Explicit extraction methods extract user features directly using interviews and questionnaires [12], which is intuitive and easy to implement; however, users may not be able to articulate their features clearly, particularly in a complex context manner. Furthermore, the method is unsuitable for large-scale applications; for example, it is impossible for a user to express their feature for thousands of different venues.

Therefore, an increasing number of researchers have focused on implicit extraction methods, which use various automated methods such as natural language analysis and data mining to extract user features from user comments and user check-in data. Depending on influencing factors, these methods can be divided into four categories: content-based feature extraction, geographical-based feature extraction, temporal-based feature extraction, and social-based feature extraction methods [13-14].

The content-based feature extraction method focuses on the analysis of content such as the user's age, job position, category of the venue, user comments on the venue, or venue photos [15-17].

Some methods have been proposed to locate the homes of the users, as a basis for calculating the distance to the venues from their homes [18-19]. The geographical-based feature extraction method is devoted to discovering the relationships between a user's check-in data and the distance from the user's home. Some researchers have conducted experiments to build various formulas, such as a power-low distribution formula or a naive Bayesian formula, to predict the probability of a venue being visited by a user at a certain distance [20–21], whereas other researchers are devoted to predicting the user's next location using historical check-in data [22-23].

Most users access different locations at different times, e.g., they tend to work in the morning and drink coffee or take a walk at night. Therefore, the temporal-based feature extraction method focuses on the time information related to the user check-in data, using various analytical methods, such as data mining or machine learning, to reveal which venues the user likes to visit within a certain time [24-26].

The social-based feature extraction method holds the idea that users share similar check-in patterns with their friends; correspondingly, users tend to make friends with those who share their features. These methods therefore use various strategies such as crawling through friend lists on user social accounts or clustering users with similar features to find other friends of a user, and apply the feature of their friends to infer their features [27–29].

Some studies combine more than one influencing factor to describe user features [30–31], which may be an interesting future research direction.

Although many methods have been proposed to describe users' features, to the best of our knowledge, few studies have focused on whether a user is suited to a feature model. In this study, we build a multiple feature model for each user and propose an algorithm to validate whether the user is suited to a feature model.

## 3 Method

### 3.1 Preliminaries

The definitions used in the method are as follows:

**Definition 1: User context set (UC)**

The USC is a set of attributes that can be used to describe the user and context involved in user activities.

$$UC=\{UC^i\} \tag{1}$$

**Definition 2: View Set (VS)**

The VS is a set of different perspectives that system analysts and managers use to observe user activity patterns according to their interests.

$$VS=\{V^j\} \tag{2}$$

**Definition 3: User context view feature set (UFS)**

The UCVFS is the set of user features from a contextual perspective determined from different perspectives under different scenarios.

$$UFS=\{UCVF^{ij}\|0=<|i|<|UC|,0=<|j|<|VS|\} \tag{3}$$

where:

$$UCVF^{ij} = \begin{pmatrix} \text{ucvf}^{ij}{}_{11} & ... & \text{ucvf}^{ij}{}_{1|V^j|} \\ \vdots & \ddots & \vdots \\ \text{ucvf}^{ij}{}_{|C^i|1} & \cdots & \text{ucvf}^{ij}{}_{|C^i||V^j|} \end{pmatrix} \tag{4}$$

### 3.2 Overview of the method

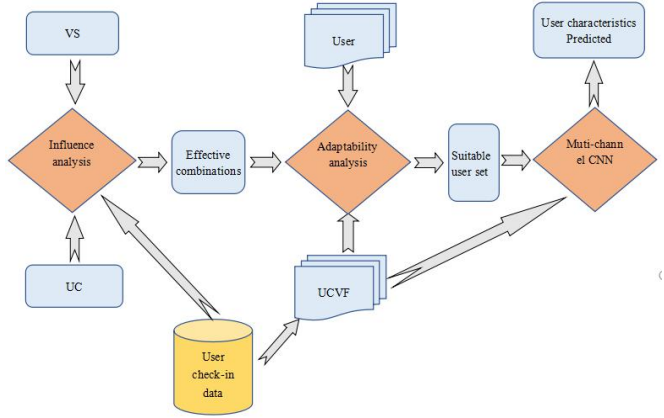

Fig. 1. Overview of the method

The overall framework of the method is shown in Fig.1. The method is based on user check-in data. Firstly, the set of view and user context is determined by the analyst, and information entropy gain is used to quantify the impact of user context to the observer's view, The specific process is described in algorithm 1 in this chapter.

Secondly, Algorithm is designed to realize the vectorization of user features, which is shown in algorithm 2. Thirdly, an algorithm based on difference value is proposed to realizes the user's applicability analysis to different features, which is shown in algorithm 3. Finally, a muti-channel CNN is designed to characterize the applicability of users to different features and used to predict user features.

This method provides a feasible process and technical route for realizing the construction and personalized analysis of user features in complex mobile applications, and has a certain promotion significance.

### 3.3 Information entropy gain based influence analysis

When the cardinality of the user context set is | UCS | and the cardinality of the view set is | VS |, dimensionality of the influence combination is calculated as shown in formula 5.

$$|ES| = |UCS| * |VS| \tag{5}$$

When the cardinality of user context set and view set increases, the cardinality of the influence set will increase rapidly, which is disadvantageous to storage and analysis. And in real life, not all user context set have a significant impact on each view. Therefore, the quantitative analysis of the impact of user context on the view is of great significance.

The concept of entropy has been widely used in thermodynamics, sociology and information science. Entropy can be used to describe the degree of chaos within the data, as shown in formula 6, where N is the number of categories and $P_i$ is the probability of belonging to category i.

$$\text{Entropy}(S) = -\sum_{i=1}^{n} p_i * \log p_i \tag{6}$$

Entropy can also be used to measure the effectiveness of data partitioning. The entropy values before and after the data partition are calculated respectively. If the entropy value is smaller after the partition, it shows that the data can be clarified through the partition, thus indicating that the partition is valuable and produces information gain, the information gain and the information gain rate are calculated as shown in formulas 7 and 8.

$$\text{Gain}(S,A) = \text{Entropy}(S) - \sum_{v \in Values} (A) \frac{S_v}{S} \text{Entropy}(S_v) \tag{7}$$

$$\text{Gain\_ratio}(S,A) = \frac{\text{Gain}(S,A)}{\text{Entropy}(S)} \tag{8}$$

In these formulas, A represents the attribute that divides the sample set S, Values represents the value set of the attribute A, v represents a value in the Values set, and $S_v$ represents the sample set corresponding to the value v after the division.

Based on the analysis above, this study uses information gain rate to find the user context that has a significant impact on the view, as shown in Algorithm 1.

Algorithm 1: influence analysis algorithm

**Input**: DS (Data Set), UCS (User Context Set), VS (View Set)

---

**Output**: ES_S (Effect Set Selected)

1: initialize ES_S=NULL

2: for each $V^j$ in VS

3:    Computes Entropy ($V^j$) based on DS using formula (6)

4:    for each $UC^i$ in UCS

5:       Computes Gain ($V^j$, $UC^i$) based on DS using formula (7)

6:       Computes Gain_ratio ($V^j$, $UC^i$) using formula (8)

7:          if (Gain_ratio ($V^j$ ,$UC^i$) >$\delta$

8:             ES_S.add (<$V^j$ ,$UC^i$>)

9:          endif

10:    endfor

11: endfor

12: return ES_S

---

The input of the algorithm includes data set, user context set and view set, and the output is the selected effect set. Firstly, initialize the selected effect set to empty(line 1), secondly, calculate the entropy of each view $V^j$ in the view set (line 2-3), and then calculate the information gain and information gain rate for each user context $UC^i$ on the view $V^j$ (lines 4-6), if the information gain rate is greater than the given threshold $\delta$, indicating that $UC^i$ has a significant influence on $V^j$, the combination <$UC^i$,$V^j$> is added to the influence set ES_S(lines 7-9).

### 3.4    UFS Construction

How to quantify user characteristics by analysing user activity data is the basis of using automatic methods to analyse user features and discover user roles. In this study, we first implement the vectorization from user activity data to user features. According to Definition 3, the UFS is a set of cardinalities | VS | * | UC |, where each set element is a matrix of |$UC^i$| times |$V^j$| dimensions. The UFS construction algorithm is presented in Algorithm 2.

---

**Algorithm 2**: UFS construction algorithm

**Input**: DS (Data Set), UC, VS

**Output**: UFS

1: Initialize data of user u DS_u from DS

2: for $UC^i$ in UC

3:    for $V^j$ in VS

4:       Initialize $UF^{ij}$ = O(|$UC^i$|*|$V^j$|)

5:       for d in DS_u

6:          uc_serial_num=0

---

```
7:        vs_serial_num=0
8:        for uc in UC^i
9:          for v in V^j
10:             if (d.UC^i_value==uc && d.V^j_value == v)
11:               uc_serial_num=GetSerial_Number(uc)
12:               vs_serial_num=GetSerial_Number(v)
13:               break
14:             endif
15:          endfor
16:          break
17:        endfor
18:        UCVF^{ij}[uc_serial_num][vs_serial_num] ++
19:      endfor
20:      UFS.add(UF^{ij})
21:   end for
22:  end for
23:  return UCVFS
```

The input of the algorithm includes the DS, UC, and VS, and the output is the UFS. First, the check-in data of user u are initialized from the DS (line 1). For each context $UC^i$ in the UC, each perspective $V^j$ in VS is iterated. For this process, $UF^{ij}$ is first initialized to a 0 matrix; the row and column are the cardinality of the $UC^i$ and $V^j$ (row 4). Iteration takes place over data record d of user U, initializing the row and column subscripts of the matrix corresponding to each record to 0 (rows 5 – 7). The corresponding user context value uc is found and the value v is viewed. The subscript of uc, i.e., uc_serial_num, and the subscript v, i.e., v_serial_num (lines 8 – 17) are calculated. The value of the matrix corresponding to the position of the row and column subscripts is added (line 18). Finally, $UF^{ij}$ is added to the UFS (line 20).

### 3.5  Applicability analysis

The previous section described the building of the UCVFS but did not analyze the applicability of the user to these features. Different users differ greatly in their perspective features. If some users of a smart location service have a regular lifestyle and a stable time for eating, working, and participating in outdoor activities every day, then the user is more suitable for the characteristics described by the time scenario and the POI category perspective. Some users do not move regularly over time but behave regularly in terms of distance. Their interest points in visiting gourmet food, for example, are generally closer to home, and most of them are within 1 km. When they visit tourist and transportation interest points, the distances are generally greater, with most concentrated at distances of more than 1 km.

In summary, it is meaningful to analyze the applicability of users to different features based on the set of user perspective features to improve the accuracy of the user descriptions. Based on the above findings, this paper proposes a method for analyzing the applicability of user-perspective features based on difference values. The method assumes that when the user applies to a particular perspective feature, the difference in the user's activity data over a long period of time is small, as shown in Algorithm 3.

---

**Algorithm 3**: Applicability analysis algorithm

**Input**: DS (Data Set), UCVFS, U(user set)

**Output**: US_UCVF$^{ij}$ (User Set_VCVF$^{ij}$, UCVF$^{ij}$ ∈ UCVFS)

1:   for each UCVF$^{ij}$ in UCVFS

2:     Initialize List SUM_UCVF$^{ij}$ =null

3:   endfor

4:   for each user u in U

5:     initialize user u's data DS_u from DS

6:     initialize distinct month M from DS_u

7:     for each m in M

8:       initialize user u's data in month m DS_um from DS_u

9:       for each UCVF$^{ij}$ in UCVFS

10:         create UCVF$^{ij}_m$ in month m using Algorithm 1

11:       endfor

12:     endfor

13:     calculate the average value of UCVF$^{ij}_m$, denote as AVG_UCVF$^{ij}_m$

14:     for each UCVF$^{ij}$ in UCVFS

15:       initialize sum_UCVF$^{ij}$ =0

16:     endfor

17:     for each UCVF$^{ij}$ in UCVFS

18:       for each m in M

19:         sum_UCVF$^{ij}$+= |UCVF$^{ij}_m$ - AVG_UCVF$^{ij}_m$

20:       endfor

21:     endfor

22:     for each UCVF$^{ij}$ in UCVFS

23:       SUM_UCVF$^{ij}$.add(u, sum_UCVF$^{ij}$)

24:     endfor

25:   endfor

26: for each UCVF$^{ij}$ in UCVFS

27:   sort SUM_UCVF$^{ij}$ order by sum_UCVF$^{ij}$

28: endfor

29: for each user u in U

30:   for each UCVF$^{ij}$ in UCVFS

```
31:        sequence_ UCVFij=GetSequence(SUM_ UCVFij,u)
32:    endfor
33:    for each UCVFij in UCVFS
34:        min_sequence=min(sequence_ UCVFij)
35:    endfor
36:    for each UCVFij in UCVFS
37:        if(min_sequence==sequence_ UCVFij)
38:            US_UCVFij.add(u)
39:        endif
40:    endfor
41: endfor
42: return US_UCVFij (User Set_UCVFij, UCVFij∈UCVFS)
```

The algorithm first initializes the |UCVFS| list, which is used to store the difference value (lines 1－3) of the user's view feature in each scenario. Through the following process, we calculate the difference value of each user's perspective feature (lines 4－25). First, we take a fixed time unit (such as the month) and establish the user's perspective feature (lines 5－12). The third step is to initialize the difference value of the user's perspective feature to 0 (lines 14－16). The fourth step is to calculate the difference value of the user's perspective feature, that is, the sum of the differences between the contextual perspective features and the mean values for each time unit (lines 17－21) and add the differences to the corresponding list (22－24). Sort the list by the difference values (lines 26－28). Calculating the order of user U in each list, and taking the smallest order corresponding to the list of scenario perspective features, when adding the user to the user set corresponding to the list of scenario perspective features, because the order is the smallest, it is demonstrated that the user is most suitable for the scenario view feature (lines 29－41) if the difference between the user and the scenario view feature is the smallest.

### 3.6  Unified model—muti-channel CNN

Multichannel neural networks can effectively describe the local saliency features of data, identify and analyze them, and then stack these different channels using a deep structure to support the fusion of multiple salient features. This feature is suitable for describing the user's adaptability to muti-perspective features; therefore, this study designs a muti-channel convolutional neural network to analyze the user's personalized features. The basic network structure is illustrated in Fig. 1.

The network contains | UCS | * | VS | channels, and the input for each channel is the user set US_UCVFij, which is suitable for the UCVFij model and the matrix UCVFij for all users. The construction of UCVFij is shown in Algorithm 1, and the identification of US_UCVFij is shown in Algorithm 2. After learning the user and the user's matrix through different channels, the network can predict the user's possible activities during a given situation.

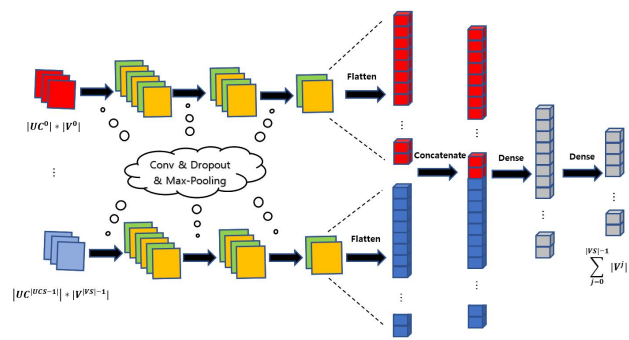

Fig. 2. Muti-channel CNN Network Structure Diagram

## 4 Experiment

### 4.1 Date sets of the Experiment

The data set is constructed from the existing Foursquare User Check-in data set [18,32]. The data set consists of three independent user check-in data sets labeled dataset_1, dataset_2, and dataset_3. Here, dataset_1 uses the New York City check-in data set from this research [32]. To verify that the method is not data sensitive, dataset_2 and dataset_3 were constructed by randomly selecting 5,000 users and 8,000 users from the data in [18]. The statistics from the data set are shown in Table 1, and example data are shown in Table 2.

**Table 1** data statistics table of the data sets

|  | User number | POI number | Check-in times |
|---|---|---|---|
| dataset_1 | 1083 | 38333 | 227428 |
| dataset_2 | 5000 | 359036 | 1472935 |
| dataset_3 | 8000 | 509440 | 2253379 |

Note: POI: Point of Interest

**Table 2** Data sample table of the data sets

| U | P | PC | PCN | LO | LA | W | Y | M | D | T |
|---|---|---|---|---|---|---|---|---|---|---|---|
| 1 | 4d* | 4b* | American Restaurant | 40.7* | -73.9* | Sat | 2012 | Apr | 07 | 17:42:24 |
| 49 | 42* | 4a* | Railway Station | 40.7* | -73.9* | Wed | 2012 | Apr | 04 | 12:11:28 |
| … | … | … | … | … | … | … | … | … | .. | … |
| 712 | 4c* | 4f* | Neighbour hood | 40.7* | -73.9* | Mon | 2012 | Nov | 05 | 23:48:22 |

Note: U: user, P: POI, PC: POI category, PCN: POI category name, LO: longitude, LA: latitude, W: week, Y: year, M: month, D: day, T: time. P and PC in the data are represented by

a string with a length of 24, for the concise and intuitive, only the first two characters are given in the table, and the rest are replaced by *. The longitude and latitude in the data are accurate to 15 decimal places after the decimal point, for the concise and intuitive, 2 decimal places are given, and the rest are replaced by *.

In the original data set, the POI category is included because the number of check-ins is limited in a particular category. To provide an intuitive understanding of user characteristics at the abstract level, according to the existing POI categories in the data, the root category is added to the data set by using the dependency relationship between the category and the root category in the category hierarchy tree on the Foursquare website. In the POI category tree of the Foursquare hierarchy, there are nine root categories: arts and entertainment, college and university, food, outdoors and recreation, professional and other places, residence, shops and services, travel and transport, and events.

### 4.2 Research Question

The research questions in this chapter are as follows:

**RQ 1**: Is it effective to assume that the user is more suitable for a specific feature model if there is little difference in $UCVF^{ij}$?

**RQ 2**: Does using the suitable $UCVF^{ij}$ help predict user behaviors.

**RQ 3**: Does the unified model improve the prediction accuracy? How does it compare with existing methods?

### 4.3 Evaluation plan

In the experiment described in this chapter, for each dataset, 80% was used as the training set, 10% was used as the verification set, and the remaining 10% was used as the test set. Because each user is created separately in $UCVF^{ij}$, the partition of the data set is also divided according to the data of each user, that is, the check-in data of each user are divided, and the union of all user check-in data is then taken as the final data set.

The top-K accuracy rate is used as an evaluation index, and the specific calculation is as shown in formula 9:

$$Accuracy @ K = \frac{| \{u, l, t, a\} \mid a \in P_{u, l, t}(K), (u, l, t, a \in TS) |}{| TS |} \quad (9)$$

In the formula, $\{u, l, t, a\}$ refers to an activity a of user u at time t at position l, and $P_{u,l,t}(K)$ refers to the top-K activity of the user at location l in time T inferred from the model. TS refers to the test set.

### 4.4 Results and Analysis

Through the analysis of the data, the user context set $UCS = \{UC^t, UC^d\}$, where t represents time and d represents distance. In this study, the time is segmented in hours, and thus $|UC^t|=24$. The distance from the user's home to the POI is divided into four levels, which are within 1 kilometre, between 1 and 10 kilometers, between 10 and 30 kilometers, and more than 30 kilometers, so $|UC^d|=4$.

According to the actual situation in the data, The view set $VS = \{V^r, V^c\}$, where r represents the root category and c represents the category of the POI. There are 9 types of POI root categories and 65 types of POI categories, and thus $|V^r|$ =9, $|V^c|$ =65.

Based on the analysis above, UCVFS = {UCVF$^{\text{time-root category}}$, UCVF$^{\text{time-category}}$, UCVF$^{\text{distance-root category}}$, UCVF$^{\text{distance-category}}$}. Among the elements above, UCVF $^{\text{time-root category}}$ is a 24*9 matrix, UCVF $^{\text{time-category}}$ is a 24*65 matrix, UCVF $^{\text{distance-root category}}$ is a 4*9 matrix, UCVF $^{\text{distance-category}}$ is a 4*65 matrix. The construction of these matrix are as shown in Algorithm 1.

After the UCVFS is constructed, Algorithm 2 is used to analyze the user's adaptation to different values of UCVF$^{ij}$. The specific results are presented in Table 3.

Table 3 Applicable users of different UCVF$^{ij}$

|  | User number | US_UCVF$^{\text{time-root category}}$ | US_UCVF$^{\text{time-category}}$ | US_UCVF$^{\text{distance-root category}}$ | US_UCVF$^{\text{distance-category}}$ |
|---|---|---|---|---|---|
| dataset_1 | 1083 | 526 | 82 | 429 | 46 |
| dataset_2 | 5000 | 1396 | 1413 | 1410 | 781 |
| dataset_3 | 8000 | 2249 | 2240 | 2258 | 1253 |

After the user's adaptability analysis of different values of UCVF$^{ij}$ is completed, the user set and matrix suitable for different values of UCVF$^{ij}$ are used as input, a multi-channel CNN is used for learning, and user activities are predicted. The specific effect is shown in detail in the result analysis of Question 3 of this section.

After the experimental results were completed, according to the research questions, the experimental results were analyzed as follows:

**RQ 1**: Is it effective to assume that the user is more suitable for a specific feature model if there is little difference in UCVF$^{ij}$?

To verify this problem, the study first calculated the difference in sum_UCVF$^{ij}$ for multiple time periods according to Algorithm 2, and then divided the difference value, starting from 10 and dividing it from 10 to 100. In the experiment, Eq. (5) was used to calculate the Top-K accuracy rates of the UCVF$^{ij}$ of user groups with different differences, where K was set to 1. The specific experimental results of the three datasets 1, 2, and 3 are shown in Fig. 3, 4, and 5, respectively.

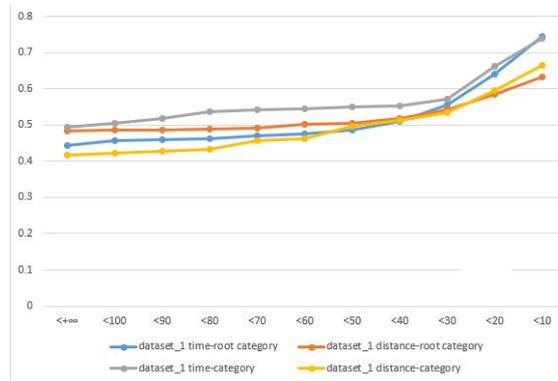

Fig. 3. Accuracy rate versus difference value change of dataset_1

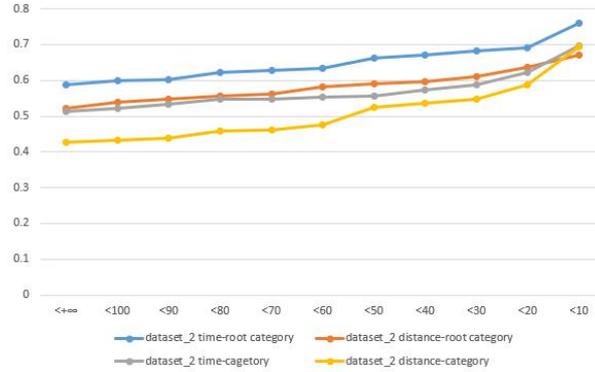

Fig. 4.   Accuracy rate versus difference value change of dataset_2

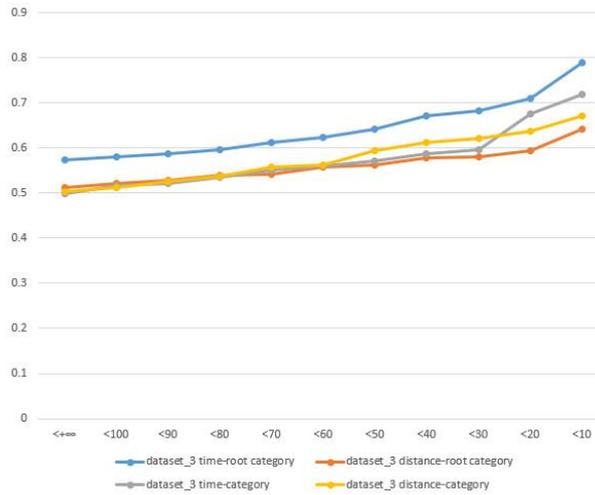

Fig. 5.   Accuracy rate versus difference value change of dataset_3

It can be seen from these figures that in all three data sets, for each of the four UCVFs, as the difference value continues to decrease, the accuracy of each UCVF increases, which indicates that when the user's long-term difference in the scene view characteristics is smaller, the assumption that the user has a higher accuracy in the feature model is valid.

**RQ 2:** Does using the suitable UCVF$^{ij}$ help predict user behaviors.

To verify this problem, this study first uses the UCVF $^{time-root\ category}$, UCVF $^{time-category}$, UCVF $^{distance-root\ category}$ and UCVF $^{distance-category}$ separately to describe all users, and calculate the accuracy rate using Eq. (5). Then, Algorithm 2 is used to divide users according to the applicability and describe them using the UCVFij applicable to the users after the division. Equation (5) is used to calculate the accuracy rate. Here, K takes the value of 1, and the experimental results are shown in Fig. 6.

It can be seen from the figure that in all three datasets, after the users are divided according to their applicability, using the user's applicable UCVF to predict their behaviors, the accuracy is higher than using each UVCF separately to predict all users. Therefore, the applicability of this user analysis helps to improve the accuracy of the user behavior prediction.

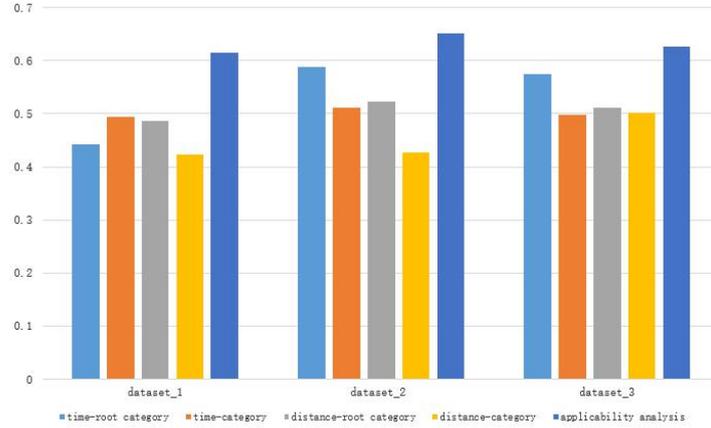

Fig. 6.   Effectiveness of applicability analysis

**RQ 3:** Does the unified model improve the prediction accuracy? How does it compare with existing methods?

- **Baseline**

In a 2020 review by Xu et al., which examined in detail the prediction of user activities in LBSN [33], the problem was categorized in terms of timeliness of prediction, and user activity prediction can be divided into the next prediction problem and any time prediction problem. The prediction of user activities can be divided into coarse-grained and fine-grained predictions based on the prediction granularity. Coarse-grained prediction includes a prediction of the POI category or a prediction of the user activity area. Fine-grained prediction refers to the prediction of user-activity POI.

According to the classification of the problem, this research belongs to any time prediction in terms of timeliness, and it belongs to coarse granularity prediction. Therefore, the baseline method of the comparison is a high-order singular vector decomposition (HOSVD), personal functional region (PFR), probabilistic category-based location recommendation (PCLR), and spatial temporal feature (STAP) [32,34-36]. The four methods were chosen as the baseline for comparison for the following reasons. First, HOSVD is a method for analyzing users from the perspective of the time series, which is often used as the baseline of the tensor decomposition method. PFR is a method for analyzing users from the active functional area, and both the PCLR and STAP methods are comprehensive methods that consider the influence of the time series and position. Second, based on the effects, these methods achieve good results for any user activity category prediction problem.

- **Effectiveness of the method**

To verify this problem, an adaptive analysis and a unified model, muti-channel CNN were compared with the four baseline methods, and the accuracy was calculated using Eq. (5). For consistency with the baseline method, the values K of top-K are 1, 5, and 10, as shown in Figs. 7, 8, and 9, respectively.

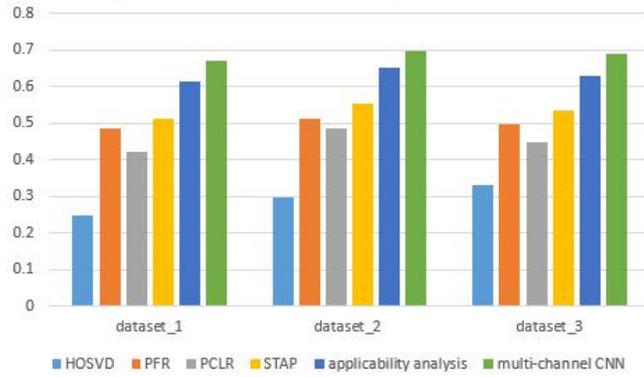

Fig. 7. Top 1 accuracy comparison

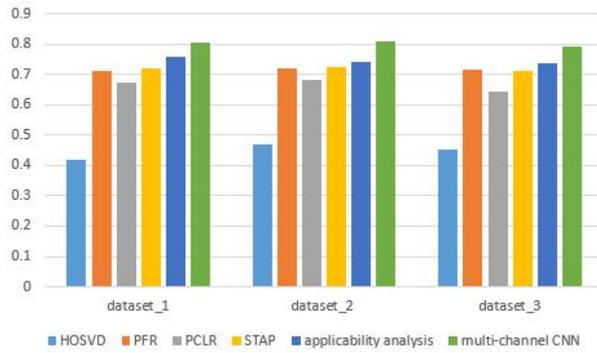

Fig. 8. Top 5 accuracy comparison

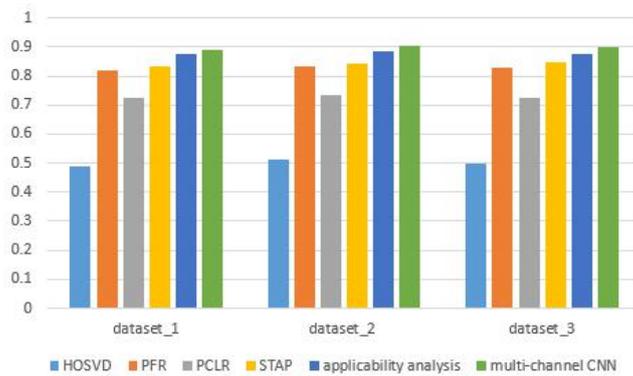

Fig. 9. Top 10 accuracy comparison

From the graph, we can see that the adaptive analysis method and the muti-channel

CNN method are better than several baseline methods for the three data sets with top-1, top-5, and top-10 results.

## 5  Validity threats

According to the criterion proposed by Wohlin et al. [37], the threat to the validity of the experiment is discussed from the following aspects:

**Conclusion validity:** In the results of the experiment, only the effect is shown, and the next step is to use statistical tests to improve the validity of the results.

**Construct validity:** The top K accuracy rate was used to analyze the experimental results. The accuracy rate is the most important index in research on intelligent services. The recall rate, F value, and other factors will be considered in a following study, and further verification of the experimental results will be carried out.

**External effectiveness:** Different data sets have different data compositions and characteristics, which may lead to changes in the effectiveness of the method. Therefore, multiple data sets from different sources were selected to verify the effectiveness of the proposed method.

## 6  Conclusions

In this study, multiple user feature models were built based on user check-in data in LBSN. Based on the feature models, an applicability analysis algorithm was proposed to find a suitable user set for a specific feature model. A unified model was used to describe the user's applicability to different feature models. Finally, experiments conducted on three data sets indicate that our method outperforms many baseline approaches.

In the future, we plan to consider the user's social attributes to construct a user feature model. In addition, the method should be validated using more data sets from different sources.

**Acknowledgments**

This work is supported by the National Natural Science Foundation of China under Grant No. U1504602.